\documentclass[12pt]{article}
\usepackage{color}
\usepackage{amsmath}
\usepackage{amssymb}
\usepackage{graphicx}
\usepackage{dcolumn}
\usepackage{bm}
\usepackage{hyperref}
\usepackage{nameref}
\usepackage{multirow,ulem}
\usepackage{array}
\usepackage{ulem}
\usepackage{authblk}
\usepackage{nameref}
\usepackage{setspace}
\usepackage[margin=1in]{geometry}

\newcolumntype{P}[1]{>{\centering\arraybackslash}p{#1}}

\title{The impact of stochastic resetting on resource allocation: The case of Reallocating geometric Brownian motion}

\author{Petar Jolakoski\textsuperscript{1}, Pece Trajanovski\textsuperscript{1,2}, Arnab Pal\textsuperscript{3,4}, Viktor Stojkoski\textsuperscript{5,6}, Ljupco Kocarev\textsuperscript{1,7}, Trifce Sandev\textsuperscript{1,2,8}}

\affil{\footnotesize
\textsuperscript{1}\footnotesize Research Center for Computer Science and Information Technologies, Macedonian Academy of Sciences and Arts, Bul. Krste Misirkov 2, 1000 Skopje, Macedonia \\
\textsuperscript{2}\footnotesize Institute of Physics, Faculty of Natural Sciences and Mathematics, Ss.~Cyril and Methodius University in Skopje, Arhimedova 3, 1000 Skopje, Macedonia \\
\textsuperscript{3}\footnotesize The Institute of Mathematical Sciences, CIT Campus, Chennai 600113 \\
\textsuperscript{4}\footnotesize Homi Bhabha National Institute, Training School Complex, Anushakti Nagar, Mumbai 400094 \\
\textsuperscript{5}\footnotesize Faculty of Economics, Ss.~Cyril and Methodius University in Skopje, 1000 Skopje, Macedonia \\
\textsuperscript{6}\footnotesize Center for Collective Learning, University of Corvinus, Budapest, Hungary \\
\textsuperscript{7}\footnotesize Faculty of Computer Science and Engineering, Ss. Cyril and Methodius University, PO Box 393, 1000 Skopje, Macedonia \\
\textsuperscript{8}\footnotesize Department of Physics, Korea University, Seoul 02841, Korea}

\date{November 18, 2024}

\begin{document}

\maketitle

\begin{abstract}
We study the effects of stochastic resetting on the Reallocating geometric Brownian motion (RGBM), an established model for resource redistribution relevant to systems such as population dynamics, evolutionary processes, economic activity, and even cosmology. The RGBM model is inherently non-stationary and non-ergodic, leading to complex resource redistribution dynamics. By introducing stochastic resetting, which periodically returns the system to a predetermined state, we examine how this mechanism modifies RGBM behavior. Our analysis uncovers distinct long-term regimes determined by the interplay between the resetting rate, the strength of resource redistribution, and standard geometric Brownian motion parameters: the drift and the noise amplitude. Notably, we identify a critical resetting rate beyond which the self-averaging time becomes effectively infinite. In this regime, the first two moments are stationary, indicating a stabilized distribution of an initially unstable, mean-repulsive process. We demonstrate that optimal resetting can effectively balance growth and redistribution, reducing inequality in the resource distribution. These findings help us understand better the management of resource dynamics in uncertain environments.
\end{abstract}




\section{Introduction}

The Reallocating Geometric Brownian Motion (RGBM) model, is a versatile framework used to understand resource redistribution governed by multiplicative dynamics across various domains~\cite{berman2020wealth,bouchaud2000wealth,liu2017correlation,marsili1998dynamical,furioli2017}. For example, in population dynamics, RGBM can represent the distribution and growth of species across different habitats, where resources such as nutrients or space are redistributed over time~\cite{robert1967,Loreau2003,levin1974,chesson2000,holt1985}. Similarly, in evolutionary biology, RGBM can model the distribution of genetic traits within a population, accounting for mutations and natural selection processes~\cite{Kimura1968EvolutionaryRA, wright1931, gillespie1994, crow1970, ewens2004}. In physics, when the RGBM dynamics are considered on a regular lattice in $d$ dimensions, upon a Cole-Hopf transformation, the model maps in the Kardar-Parisi-Zhang (KPZ) equation~\cite{gueudre2014explore} with applications in a range of areas: cosmology and turbulence \cite{frisch2001new, bec2007burgers}, surface growth \cite{krug1991kinetic, barabasi1995fractal} and directed polymers \cite{halpin1995kinetic}. In econophysics, RGBM is also known as the Bouchaud-Mézard model of wealth condensation~\cite{bouchaud2000wealth}, and is used to simulate the allocation of wealth among individuals or the distribution of resources across different sectors of an economy.

Depending on the parameters, RGBM can lead to either positive or negative reallocation of resources. Positive reallocation, where resources flow from the rich to the poor, results in a system that is stationary, self-averaging, and ergodic~\cite{van2011stochastic, Birkhoff1931, Cohen_1999}. In this regime, the long-term behavior of the system is predictable, and the time-average matches the ensemble average, making it a stable representation of resource dynamics. In the case of negative reallocation, however, the resources are disproportionately redistributed from the poor to the rich and the system becomes unstable and non-ergodic~\cite{bouchaud2000wealth,Peters_Klein2013,ruelle1981,barkai_metzler_livingcells,bouchaud_weak_ergodicity,Barkai_aging_CTRW2003,stojkoski2022ergodicity,bouchaud_potters_financial_risk}.
In this non-ergodic regime, the time-average does not converge to the ensemble average, leading to unpredictable and often inequitable outcomes. Notably, recent literature suggests that the non-ergodic regime may be a more realistic representation of real-world resource dynamics, where wealth and resources tend to concentrate rather than distribute evenly~\cite{berman2020wealth}.

To address the challenges posed by the non-ergodic nature of RGBM in the negative reallocation regime, we explore the application of stochastic resetting~\cite{Evans2011,Evans_2011,Evans_2020,pal2015diffusion,Bartumeus_2009,pal_Kusmierz2020,Reuveni2014,pal2017first,bell2012searching,Pal_2016,Kusmierz2019,tucci2020,Stojkoski_2022,Vinod2022,stojkovski_income_inequality,Christophorov_2022,bonomo_pal2021,riascos_boyer2020,jolakoski2023first,huang2021,rose2018,Yin_quantum_restart,TalFriedman2020,Besga_Benjamin,trajanovskiPRE,olsen2023steady,trajanovski3DCOMB,kumar2023universal,Campos2015phase,bonomo2022mitigating,roy2024queues,singh2020resetting}. Stochastic resetting involves periodically reverting the system to a predetermined state, which can potentially stabilize the dynamics of the process~\cite{Evans2011}. In real-world applications, resetting can represent various interventions depending on the domain. In population dynamics, resetting might correspond to natural events or human interventions, such as habitat restoration or reintroduction of species, which replenish resources or reset population levels. In evolutionary biology, it can represent environmental changes that disrupt selection pressures or reintroduce genetic diversity through migration or mutation. In physics, resetting could correspond to external forces or events that disrupt surface growth or turbulence, such as cooling processes in surface growth or shocks in cosmology. These interventions allow systems to avoid collapse, re-stabilize, or explore new configurations. See \cite{Evans_2020,pal2023random,gupta2022stochastic} for a multi-facet review of the subject. 

\begin{figure}
\centering
\includegraphics[width=14cm]{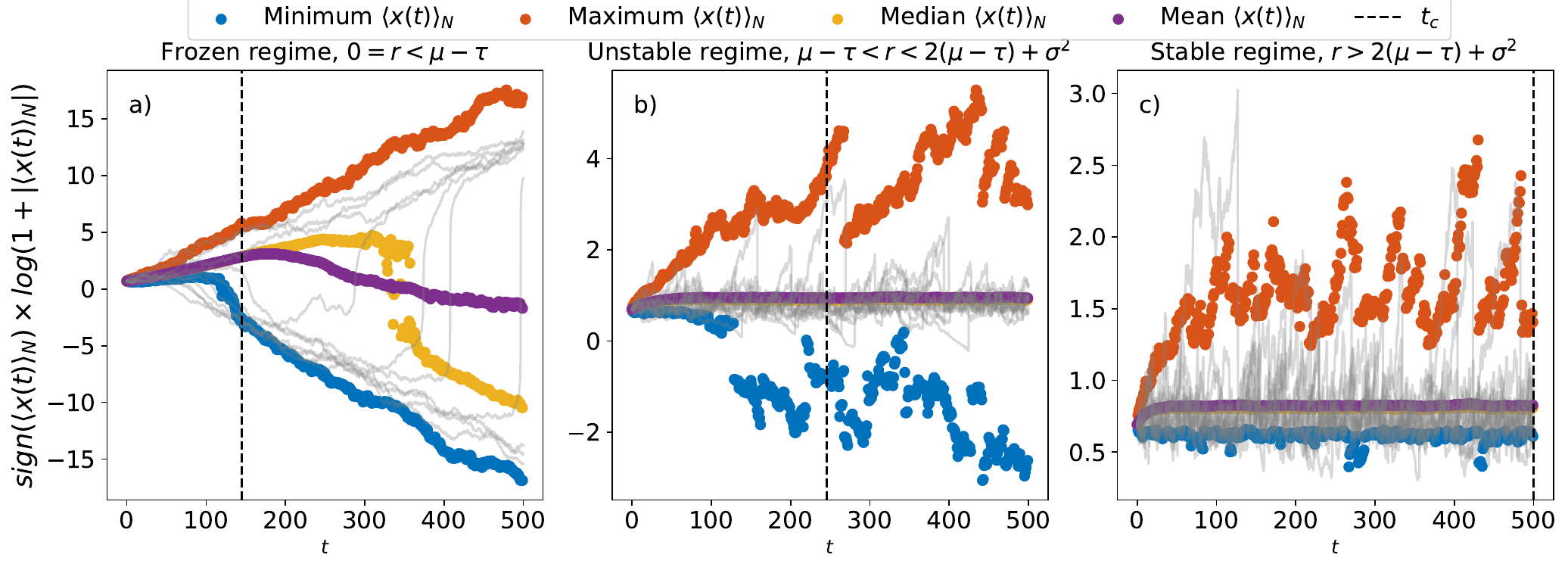}
\caption{\textbf{Mean wealth behavior in different regimes.} Numerical estimations for the median, mean, maximum and minimum of the mean wealth $\langle x(t) \rangle_N$ for a super sample of $10^3$ simulations with $N=10, \mu=0.021, \sigma^2=0.01, \tau=-0.01$. a) No resetting, $r=0$, b) $\mu-\tau<r<2(\mu-\tau)+\sigma^2$ and c) $r>2(\mu-\tau)+\sigma^2$. $t_c$ is critical self-averaging time, which for c) is practically infinite. The gray lines in the background are samples of trajectories of a single simulation run.}
\label{fig:pdf_trajs_}
\end{figure}

We find three different long-term regimes specific to the negative reallocation regime. These regimes depend on the relation between the resetting frequency, the magnitude of the reallocation and standard GBM parameters: the drift and the noise amplitude. In the first regime, the resetting rate is insufficient to mitigate the mean-repulsive behavior, resulting in dynamics similar to the standard RGBM (see  Fig.~\ref{fig:pdf_trajs_}a). As the resetting rate increases, however, the system transitions into the second and third regime, where the first and second moments converge, respectively. The behavior of the mean in these regimes is visualized in subplots b) and c) in Fig.~\ref{fig:pdf_trajs_}. The convergence of the both moments in the last regime hints at a stabilized probability distribution of the initially unstable mean-repulsive process, which is one of the main results of this work, because it enables us to explore the system's dynamics and derive analytical results under the condition of negative resource reallocation, an aspect that has not been addressed in previous studies. 
Moreover, in the third regime, the self-averaging time period becomes practically infinite and in this case the system will mimic ergodic behavior.

To demonstrate implications of our results, we consider the particular case of wealth redistribution and the effects of resetting on mobility and wealth concentration, assuming that they undergo the RGBM with resetting process. In \cite{stojkoski2024measures} the authors study the measure of mixing in standard RGBM in order to quantify mobility across the whole wealth distribution. They argue that the relationship between standard mobility measures and mixing in the negative reallocation regime is impossible. Additionally, they show that some mobility measures such as Spearman's rank correlation, the Intragenerational earnings elasticity (IGE) and transition matrices will still indicate a presence of mobility in the non-mixing regime (negative reallocation). Their results for Spearman's correlation and IGE suggest that in the negative reallocation regime, the magnitude of the reallocation rate does not impact the extent of mobility. Here we recover the same result that the rank correlation remains unaffected by the magnitude of the negative reallocation rate and decreases as a function of the resetting rate. Moreover, there is a weak dependence between earnings elasticity and the strength of negative reallocation at lower resetting rates. However, as the resetting rate increases, earnings elasticity converges to a constant value, independent of reallocation strength. Finally, we study the degree of concentration with the probability of observing states located in the top 1\% of the distribution as is done in \cite{stojkoski2021geometric} and recover the same result that as the resetting rate increases, the probability of observing extreme configurations decreases and stabilizes.

These results help us understand how stochastic resetting can mitigate the inherent instability of the non-ergodic regime and create conditions that are more favorable for equitable resource redistribution. 

The remainder of this paper is structured as follows. In Section \ref{sec:model}. we provide a detailed description of the Reallocating Geometric Brownian Motion (RGBM) model with stochastic resetting along with its jump-diffusion formulation. Section \ref{sec:theoretical-framework}. develops the theoretical framework for the analytical results using the jump-diffusion formulation. Section \ref{sec:results}. delves into the analysis of the effects of stochastic resetting on RGBM, highlighting the identification of the three distinct long-term regimes and exploring their implications for wealth redistribution. Finally, in Section \ref{sec:discussion}. we summarize our findings, discuss their broader implications and limitations; and finally propose potential avenues for future research.

\section{Model} \label{sec:model}

In our model, at each time point $t$, the resources $x_i(t)$ of entity $i$ evolve through one of two processes. First, with probability $1 - rdt$, the resources grow multiplicatively, following a growth rate $\mu$ and experiencing stochastic fluctuations with noise amplitude $\sigma$. Alternatively, with probability $rdt$, the resources are reset to a fixed value $x_r$. After the growth or reset phase, each entity contributes a fraction $\tau$ of their resources into a central pool. The total amount in the pool is then redistributed evenly across all entities in the population, ensuring that every individual receives the same share of the collective resources, regardless of their initial contribution.

The model is usually defined with the following Langevin equation:
\begin{align}\label{RGBM_resetting_langevin}
    dx(t)=& \, (1-Z_t) \left[x(t)(\mu dt + \sigma dW_t) - \tau (x(t)-\langle x \rangle_N)dt\right] + (x_r - x(t))Z_t,
\end{align}
where $\mu$ is the drift term, $\sigma$ is the noise amplitude, $dW_t$ is an independent Wiener increment, $\langle x \rangle_N $ is the ensemble or population average and $\tau$ is the rate of reallocation. Resetting events occur when the random variable $Z_t$ takes value 1 in the interval between $t$ and $t+dt$; otherwise, it is zero and the dynamics correspond to the standard RGBM. The corresponding Fokker-Planck equation for the model~(Eq.~\ref{RGBM_resetting_langevin}), in the It\^{o} interpretation of the multiplicative noise $\eta(t)=dW_t/dt$, by utilising~\cite{risken1996fokker} takes the form:
\begin{align}\label{fokker_planck_eq_resett}
    \dfrac{\partial P(x,t)}{\partial t} =&-\frac{\partial}{ \partial x} \biggr{\{} \Big[\mu x-\tau \big(x - \langle x \rangle_N\big)\Big] P(x,t) \biggr{\}} +\frac{\sigma^2}{2} \frac{\partial^2}{ \partial x^2} \biggr{[}x^2 P(x,t)\biggr{]}-rP(x,t)+r\delta (x-x_0).
\end{align}

The model without resetting simplifies to the well-known Reallocating Geometric Brownian Motion (RGBM) equation, which models resource redistribution under multiplicative dynamics. The behavior of this system is governed by the sign of the reallocation rate  $\tau$. For $\tau > 0$, the growth rate of the population’s average resources becomes an ergodic observable, and the model exhibits mean-reverting dynamics, where each $x_i$ eventually converges to the population average. In this regime, the large population approximation for the average resources, $\langle x(t) \rangle_N = \exp{(\mu t)}$, remains valid. In contrast, when $\tau < 0$, the system becomes non-ergodic, and its dynamics bifurcate into two distinct phases, divided by a critical self-averaging time $t_c$. During the initial period, when $t < t_c$, the ensemble average approximates the population behavior, with $\langle x(t) \rangle_N \sim \exp{(\mu t)}$. However, after this time, the non-ergodicity dominates, causing the population average to be influenced by extreme values resulting in both positive and negative outcomes, due to the presence of individuals with negative resources. This phenomenon is visualized in subplot a) in Fig.~\ref{fig:pdf_trajs_} where we plot the mean wealth behavior averaged across $10^3$ simulations. We can see that after the self-averaging period non-ergodicity causes the average to oscillate between positive and negative values, as its magnitude becomes influenced by the most extreme wealth values in the population. The properties of standard RGBM are summarized in Appendix~\ref{app:prop-rgbm}.

In our case (when $r > 0$), the model incorporates resource redistribution with resetting, where periodically the resources of entities are reset to a baseline value $x_r$, which we set equal to the initial position $x_r=x_0=1$. In many scenarios, resetting may make the model more realistic. For example, in economic systems, periodic resets can model tax reforms or wealth redistribution efforts that prevent extreme inequalities. In ecological systems, resetting could represent natural events like wildfires or floods that reset populations or resource availability. Similarly, in financial markets, resets could model regulatory interventions or market corrections.

\subsection{Jump-diffusion formulation}

We proceed with an alternative formulation of the process (Eq.~\ref{RGBM_resetting_langevin}) (as done in \cite{Magdziarz2023}) which allows us to develop a framework for analytical calculation of the moments and regimes. This formulation is within the framework of the so-called jump-diffusion models, such as the L\'evy processes. Examples of these kind of processes are the $\alpha$- stable, Linnik, Mittag-Leffler, Gamma or Laplace processes~\cite{s.Ken-Iti1999}. These processes are implemented in various physical systems ~\cite{Einstein1905annalenderphysik,smoluchowski1906annalenderphysik}, financial systems~\cite{Bachelier1900,Merton1976, Black-Scholes1973}, as well in models in biology chemistry, data mining, statistics and other various fields~\cite{Barndorff2001}. Another L\'evy process that is of our interest is the Poisson process which, is a non-decreasing L\'evy process with jumps of size 1 and flat periods between jumps. Times between consecutive jumps of the Poisson process are independent and drawn from an exponential distribution with mean $r>0$. The process of interest in this paper~(Eq.~\ref{RGBM_resetting_langevin}) formulated as a jump-diffusion model of Poisson type is given with the stochastic equation~\cite{Magdziarz2023,hanson2007applied}:
\begin{align}\label{stochastic_eq1}
    dX_t=dR_t + (x_r - X_t)dN_t, \,\,\,\, x_r = x_0,
\end{align}
where $N_t$ is a Poisson process with intensity $r$, with resetting position $x_r=x_0$ and the RGBM process as $R_t$~(Eq.~\ref{eq:rgbm_process}). This form of the stochastic equation is an analogue of the stochastic equation defined in~(Eq.~\ref{RGBM_resetting_langevin}) where $x(t)=X_t$. Here $N_t=\max\{ n \in  \mathbb{N} : \sum_{i=1}^n T_i \leq t \}$ and whether there is a change in values or not, $dN_t$ can take values $1$ and $0$. When the Poisson process takes value $1$ a resetting event occurs and $(x_r - X_t)dN_t$ sends the particle to the initial position, and when $dN_t=0$ the RGBM process is active and $X_t = R_t$. This form allows us to use the stochastic It$\hat{o}$ integral~\cite{Ito1944,Ito1946} and It$\hat{o}$'s lemma in order to analyse the process. The process of interest in this paper~(Eq.~\ref{stochastic_eq1}) has the full form:
\begin{align}\label{stochastic_eq2}
    dX_t=X_t(\mu dt + \sigma dW_t) - \tau (X_t-\langle x \rangle_N)dt + (x_r - X_t)dN_t,
\end{align}

From Eq.~\ref{stochastic_eq2} we observe the two opposing forces: when $\tau<0$ the second term of the RHS acts as a repulsive force, whereas when there is a resetting event the last term of the RHS has the effect of a potential that instantaneously reverts the particle to the resetting position $x_r$. In Sec.~\ref{sec:results}, we explore in more detail this interplay between the magnitude of negative $\tau$ and the resetting rate, in relation to the GBM parameters. 

We will next develop the theoretical framework using the jump-diffusion formulation.

\section{Theoretical framework} \label{sec:theoretical-framework}

We can derive the formulation in a more general setting by taking an arbitrary function $f(X_t)$, such that using the Taylor expansion we get 
\begin{align}\label{itolemmaeq}
    df(X_t)=& \, \frac{\partial f}{\partial t}+\sum_i \frac{\partial f}{\partial X_{t,i}} dX_{t,i} + \frac{1}{2} \sum_i \sum_j \frac{\partial^2 f}{\partial X_{t,i} \partial X_{t,j}} dX_{t,i} dX_{t,j}.
\end{align}

We are interested in calculating the expected value of that arbitrary function $f(X_t)$. From~(\ref{itolemmaeq}) the general It$\normalfont{\hat{o}}$ formula for the expected value $\mathbb{E}[f(X_t)]$ is (see 4.2 in \cite{oksendal2013stochastic})
\begin{align}\label{itolemmaeq2}
    d\mathbb{E}[f(X_t)]=& \sum_i \mathbb{E}
    \biggr{[}\frac{ \partial f}{\partial X_{t,i}} dX_{t,i} \biggr{]} +\frac{1}{2} \sum_i \sum_j \mathbb{E} \biggr{[}\frac{\partial^2 f}{\partial X_{t,i} \partial X_{t,j}} dX_{t,i} dX_{t,j} \biggr{]}.
\end{align}

Now, from \cite{stojkoski2022ergodicity} and \cite{Magdziarz2023}, if we take $i=j$, and substitute the stochastic equation~(\ref{stochastic_eq2}) for $dX_{t,\{i,j\}}$ in the Taylor expansion~(\ref{itolemmaeq2}),
\begin{align}\label{itolemmaeqqq}
    d\mathbb{E}[f(X_t)]=& \sum_i \mathbb{E}
    \biggr{[}\frac{ \partial f}{\partial X_{t,i}} \biggr{(}X_t(\mu dt + \sigma dW_t) - \tau (X_t-\langle x \rangle_N)dt + (x_r - X_t)dN_t \biggr{)} \biggr{]} \nonumber\\&+\frac{1}{2} \sum_i \mathbb{E} \biggr{[}\frac{\partial^2 f}{\partial X_{t,i}^2} \biggr{(}X_t(\mu dt + \sigma dW_t) - \tau (X_t-\langle x \rangle_N)dt + (x_r - X_t)dN_t\biggr{)}^2 \biggr{]}.
\end{align}
Expanding the last term in Eq.~\ref{itolemmaeqqq}, substituting $(dW_t)^2=dt$, $(dW_t)^n =0$ for $n>2$, and using $(dN_t)^n=dN_t$ for any $n \in \mathbb{N}$ and dividing by $dt$, we end up with the following equation
\begin{align}\label{itodiffequation}
   \frac{d\mathbb{E}[f(X_t)]}{dt} = & \, (\mu-\tau) \sum_i \mathbb{E} \biggr{[}  \frac{\partial f}{\partial X_{t,i}} X_{t,i} \biggr{]}+\tau \mathbb{E} \biggr{[} \langle x \rangle_N \sum_i \frac{\partial f}{\partial X_{t,i}} \biggr{]}  \nonumber\\&+\frac{\sigma^2}{2} \sum_i \mathbb{E} \biggr{[} \frac{\partial^2 f}{\partial X_{t,i}^2}  X_{t,i}^2 \biggr{]} + r \mathbb{E}[f(x_r)-f(X_{t,i})],
\end{align}
where we have used 
\begin{align}
    &\mathbb{E}\biggr{[}f(x_r)-f(X_t)\biggr{]}dN_t \nonumber\\& =\sum_{i=0}^{1}\mathbb{E}\biggr{[}(f(x_r)-f(X_t))dN_t|dN_t=i\biggr{]}\mathbb{P}(dN_t=i)\nonumber\\& = \mathbb{E}\biggr{[}(f(x_r)-f(X_t))dN_t|dN_t=1\biggr{]}\mathbb{P}(dN_t=1) \nonumber\\& = rdt \mathbb{E}\biggr{[}f(x_r)-f(X_{t,i}) \biggr{]}
\end{align}

Now if we take $f(X_t)=x$ we can calculate the ensemble average. The solution for this is straightforward, by substituting $f(X_t)=x$ in~(\ref{itodiffequation}) and solving for $\mathbb{E}[x]$ we get the expression
\begin{align}\label{expected mean}
    \mathbb{E}[x_i(t)]=\mathbb{E}[x_0] \biggr{[}  \frac{\mu}{\mu-r} \exp{((\mu-r)t)} - \frac{r}{\mu -r}   \biggr{]}
\end{align}
where we utilized 
\begin{align}\label{self_averging_property}
     \mathbb{E}[f(x)]= \lim_{N\to\infty}  = \langle f(x) \rangle_N 
\end{align}

However, due to the non-ergodic nature of the negative reallocation regime of RGBM, Eq.~\ref{self_averging_property} will be valid until some critical self-averaging time, $t_c$, but with the introduction of stochastic resetting, for some value of the resetting rate $r_c$, the assumption will be applicable always ($t_c \rightarrow \infty$). The calculation of $r_c$ is the focus of the next section which will play a key role in calculating the moments.

\subsection{Critical self-averaging time}

To explore some properties of RGBM with resetting under negative wealth reallocation rates, we consider the concept of self-averaging. In statistical physics self-averaging is a property of the system when, a sample mean resembles the corresponding expectation value, i.e. when~Eq.~\ref{self_averging_property} is valid. The key question we ask in this paper is: What resetting rate will cause the self-averaging time to become infinite? In other words, which value of the resetting rate, $r$, will allow us to remain in the self-averaging regime indefinitely? To estimate when this occurs, we need to investigate the relative variance \cite{lobejko2018self,peters2018ergodicity} of the population average $\langle x(t) \rangle_N$, defined as
\begin{align}\label{variance_of_x}
    R_N(t)\equiv \frac{var(\langle x(t) \rangle_N)}{\langle \langle x(t) \rangle _N \rangle^2},
\end{align}
where $var(x) = \mathbb{E}[x^2] - \mathbb{E}[x]^2$ is the variance of x.

Using this value, we can identify when the system exhibits self-averaging and the population average wealth will always resemble the ensemble average; this will be true when $R_N(t)$ converges to 0 in the time limit. For negative values of the reallocation rate $\tau$ in RGBM the system will experience self-averaging until some critical time $t_c$, which is dependent on both the initial condition and the population size $N$, and afterwards it will collapse to its time-average behavior. Stochastic resetting effectively allows this critical time to become infinite for certain values of the rate $r$. The critical time can be found by writing for~(Eq.~\ref{variance_of_x})
\begin{align}
    R_N(t) &= \frac{\mathbb{E}[\langle x(t) \rangle_N^2] - \mathbb{E}[\langle x(t) \rangle_N]^2}{\mathbb{E}[\langle x(t) \rangle_N]^2}\nonumber\\& = \frac{1}{N^2} \frac{\mathbb{E}[\sum_i \sum_j x_i(t)x_j(t)]}{\mathbb{E}[\langle x(t) \rangle_N]^2}-1  \nonumber\\& = \frac{\sum_i \sum_j \mathbb{E} [x_i(t) x_j(t)]}{(\sum_i \mathbb{E}[x_i(t)])^2} - 1,
    \label{eq:relative-var}
\end{align}
where we have used the fact that $\mathbb{E}[\langle x(t) \rangle_N^2]=\frac{1}{N^2} \mathbb{E}[\sum_i \sum_j x_i(t) x_j(t)]$.
The system will be self-averaging until the critical point, which occurs when $R_N(t_c)=1$. In order to calculate the relative variance we need to estimate the dynamics of $\mathbb{E}[x_i(t) x_j(t)]$, and because any two trajectories are coupled and their evolution is interdependent when $\tau \neq 0 $, this task is not trivial. This can be done by interpreting the dynamics of $\mathbb{E}[x_i(t) x_j (t)]$ as a system of differential equations and utilizing the It$\normalfont{\hat{o}}$ lemma~\cite{stojkoski2022ergodicity} of RGBM with resetting. By setting $f(x)=x_ix_j$ in~Eq.~\ref{itodiffequation}, we see that the dynamics of $d\mathbb{E}[x_i(t) x_j (t)] / dt$ can be described as~\cite{stojkoski2022ergodicity,peters2018ergodicity}:
\begin{align}
    \begin{cases}
      2 \biggr{(} \mu-\frac{N-1}{N} \tau +\frac{\sigma^2}{2} -r \biggr{)} \mathbb{E} [x_i^2]+\frac{\tau}{N} (\sum_{k \neq i} \mathbb{E}[x_ix_k] + \sum_{k\neq i}\mathbb{E}[x_i x_k] )+r \mathbb{E}[x_0^2] \,\,\,\,\,\,\,\,\, \text{if $i=j$}\\ 
      2 \biggr{(} \mu-\frac{N-1}{N} \tau -r \biggr{)} \mathbb{E} [x_i x_j]+\frac{\tau}{N} (\sum_{k \neq i} \mathbb{E}[x_kx_i] + \sum_{k\neq j}\mathbb{E}[x_k x_i] )+r \mathbb{E}[x_0^2] \,\,\,\,\,\,\,\,\,\text{otherwise}
    \end{cases}\,
\end{align}
where we can rewrite them with $v(t)=\mathbb{E}[x_i^2(t)]$ and $ q(t)=\mathbb{E}[x_kx_j]$, so we end up with another form for the equations:
\begin{align}
    \begin{cases}
    \frac{dv}{dt} = 2\biggr{(}   \mu-\frac{N-1}{N}\tau +\frac{\sigma^2}{2} -r  \biggr{)} v + 2\frac{N-1}{N}\tau q +rv(0) \\
    \frac{dq}{dt} = 2\biggr{(}   \mu-\frac{1}{N}\tau -r  \biggr{)} q + \frac{2}{N}\tau v + rv(0)
    \end{cases}
\label{eq:system-of-eqs}
\end{align}

\begin{figure}[ht!]
\centering
\includegraphics[width=13cm]{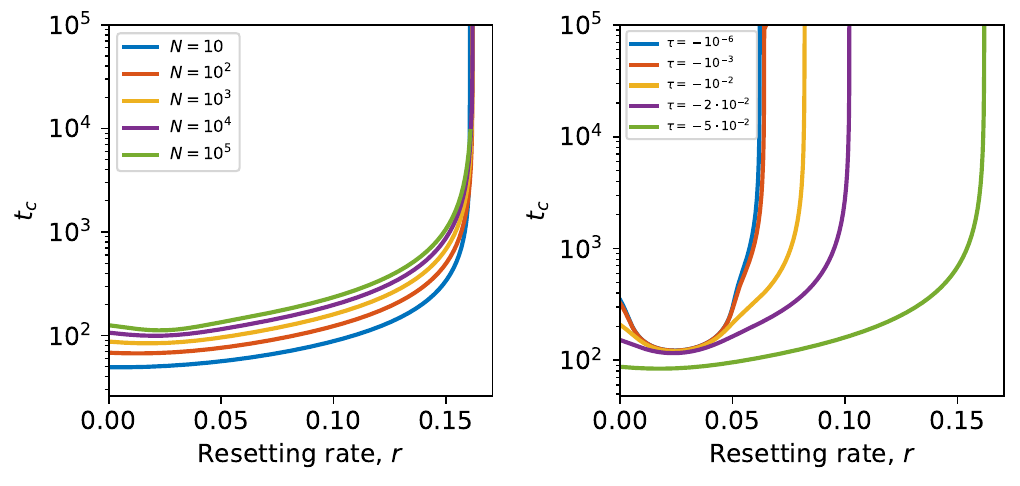}
\caption{Self-averaging critical time as a function of the resetting rate, for different number of trajectories with $\tau=-0.05$ (left), and reallocation rates with $N=1000$ (right). Parameters: $\sigma=\sqrt{0.02}$, $\mu=0.021$.}
\label{fig:critical_times}
\end{figure}

This system of equations is solvable, and we get the expressions for $q(t)$ and $v(t)$. On the other hand, the relative variance (\ref{eq:relative-var}) can be rewritten in the following form (see \cite{stojkoski2022ergodicity})
\begin{align}
    R_N(t)=\frac{v(t)+(N-1)q(t)}{N (\mathbb{E}[x_i(t)])^2}-1,
\end{align}
and by utilizing $q(t)$, $v(t)$ and $\mathbb{E}[x_i(t)]$ (\ref{expected mean}) we can calculate the resetting rates for which $t_c \rightarrow \infty$. Generally, the behavior of the critical self-averaging time, $t_c$, for various population sizes and magnitudes of negative reallocation, as a function of the resetting rate, is shown in Fig.~\ref{fig:critical_times}. The system's behavior is more strongly influenced by the value of $\tau$, compared to the impact of the population size. As $\tau$ becomes more negative, the redistribution dynamics are more unstable, requiring a higher resetting frequency to reach self-averaging.

Moreover, identical to the unstable regime (when $\mu<r<2\mu+\sigma^2$) in the baseline model (without reallocation), the ensemble average of our model (with reallocation)~(\ref{expected mean}) converges to a stationary value,
\begin{align}
   \mathbb{E} [x_i(t)]\approx \frac{r}{r-\mu} x_0,
\end{align}
and as we saw previously, self-averaging is present when the following expression is true:
\begin{align}\label{ensamble_average}
    \mathbb{E}[x_i(t)] = \langle x(t) \rangle_N.
\end{align}

We will next explore how the assumption (Eq.~\ref{ensamble_average}) is key in the analytical calculations through which we analyze the influence of resetting on the overall dynamics of the system.

\section{Results}\label{sec:results}

\subsection{Moments and classification of regimes}

In the self-averaging regime, when Eq.~\ref{ensamble_average} is true, the Fokker-Planck equation~(\ref{fokker_planck_eq_resett}) has the following form:
\begin{align}\label{Fokker-Planck_approx_in third _regime}
    \dfrac{\partial P(x,t)}{\partial t} =& \, -(\mu-\tau)\frac{\partial}{ \partial x} \biggr{[} (x+\frac{\tau r}{(r-\mu)(\mu-\tau)} x_0 )P(x,t) \biggr{]} \nonumber \\& +\frac{\sigma^2}{2} \frac{\partial^2}{ \partial x^2} [x^2 P(x,t)]-rP(x,t)+r\delta (x-x_0),
\end{align}
where we have substituted 
\begin{align}
    \langle x(t) \rangle_N=\frac{r}{r-\mu} x_0.
\end{align}
This form of the Fokker-Planck equation will enable us to calculate some properties of the RGBM with resetting model, such as the first and second moment. If we multiply Eq.~\ref{Fokker-Planck_approx_in third _regime} by $x$ and $x^2$ and integrate from $-\infty$ to $\infty$ we can calculate the mean value $\langle x(t) \rangle$ and the MSD $\langle x^2(t) \rangle $, respectively:
\begin{align}\label{mean_value_analytical}
    \langle x(t) \rangle=\frac{x_0}{r-\mu} [r-\exp{(-t(r-\mu+\tau))}\mu]
\end{align}
and 
\begin{equation}
\begin{gathered}\label{msd_analytical}
    \langle x^2(t) \rangle=\frac{r x_o^2}{(r-\mu)^2} \exp{(t(2\mu-r+\sigma^2-2\tau))} \\ \times\left[ \frac{2\exp{(-t(\mu+\sigma^2-\tau))}\mu \tau}{\mu+\sigma^2-\tau} + \frac{\exp{(t(r-2\mu-\sigma^2+2\tau))}(r^2-2r\mu +\mu ^2 +2r \tau)}{r-2\mu -\sigma^2 +2\tau}\right] \\ + \exp{(t(2\mu-r+\sigma^2-2 \tau))} \left[ x_0^2-   \frac{rx_0^2 (\frac{2\mu \tau}{\mu+\sigma^2-\tau}+\frac{r^2-2r\mu+\mu^2+2r\tau}{r-2\mu-\sigma^2+2\tau})}{(r-\mu)^2}   \right]
\end{gathered}
\end{equation}

To ensure the long-time limit of the mean and MSD converge, the conditions $r>\mu-\tau$ and $r>2(\mu-\tau) +\sigma^2$, respectively, must be met. This gives rise to three regimes for the evolution of the moments, summarized in Table~\ref{tab:regimes}. For the case without reallocation (when $\tau=0$), we recover the results for the regimes of srGBM analyzed in \cite{stojkoski2021geometric}. This simple modification of the regimes by the reallocation parameter $\tau<0$ is an important and interesting result.

\begin{table}
\centering
\caption{Moments behavior of RGBM with resetting}
\begin{tabular}{|l|ccc|}
\hline
\multicolumn{1}{|c|}{\multirow{2}{*}{\textbf{Moment}}} & \multicolumn{3}{c|}{\textbf{Limiting behavior}}                                \\ \cline{2-4} 
\multicolumn{1}{|c|}{} & \multicolumn{1}{P{2.5cm}|}{\textbf{Exponential \newline divergence}} & \multicolumn{1}{P{2.5cm}|}{\textbf{Linear \newline divergence}} & \textbf{Convergence} \\ \hline
\textbf{$\langle x(t) \rangle$}                                    & \multicolumn{1}{c|}{$r<\mu-\tau$}         & \multicolumn{1}{c|}{$r=\mu-\tau$}         & $r>\mu-\tau$         \\ \hline
\textbf{$\langle x^2(t) \rangle$}                                    & \multicolumn{1}{c|}{$r<2(\mu-\tau)+\sigma^2$}         & \multicolumn{1}{c|}{$r=2(\mu-\tau)+\sigma^2$}         & $r>2(\mu-\tau)+\sigma^2$         \\ \hline
\end{tabular}
\label{tab:regimes}
\end{table}

By finding the limit $t \rightarrow \infty$ of~equations \ref{mean_value_analytical} and \ref{msd_analytical} we get the stationary values for the first moment,
\begin{align}\label{long_mean}
    \langle x(t) \rangle_{st} = \frac{r}{r-\mu} x_0,
\end{align}
and MSD
\begin{align}\label{long_msd}
    \langle x^2 (t) \rangle_{st}=\frac{r}{(r-\mu)^2} \frac{r^2-2r(\mu-\tau)+\mu^2}{r-2(\mu-\tau)-\sigma^2} x_0^2,
\end{align}
where for $\tau=0$, the results in~\cite{stojkoski2021geometric} are again recovered.

\begin{figure}[]
\centering
\includegraphics[width=13cm]{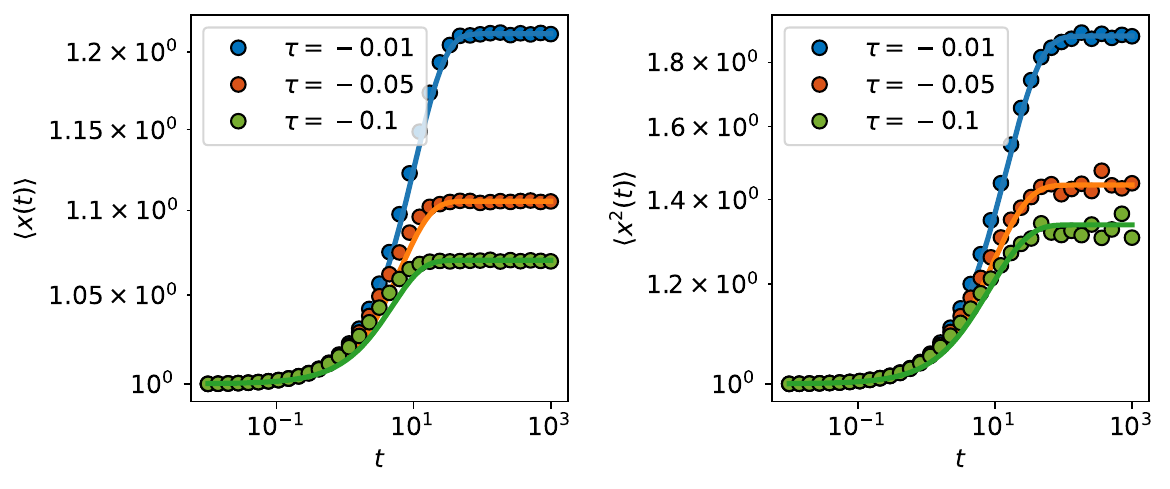}
\caption{\textbf{MEAN and MSD}. Parameters: $\mu=0.021, \sigma=\sqrt{0.01}, dt=0.01$. The resetting rates for $\tau=\{-0.01,-0.05,-0.1\}$ plots are $r=\{0.12, 0.22, 0.32\}$ respectively, chosen to ensure the convergence of both moments. The results are the average of $10^3$ simulation runs.}
\label{fig:mean-msd}
\end{figure}

\begin{figure}[ht!]
\centering
\includegraphics[width=7cm]{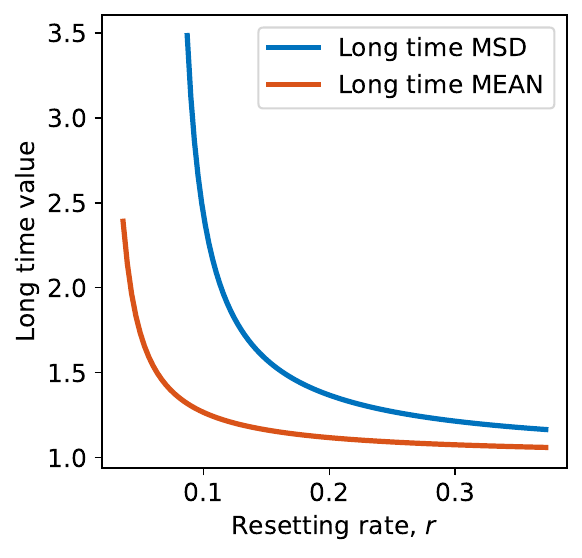}
\caption{\textbf{Long time MEAN and MSD}. Parameters: $\mu=0.021, \sigma=\sqrt{0.01}, \tau = -0.01, T=10^4$.}
\label{fig:lt-mean-msd}
\end{figure}

This long-term behavior as a function of the resetting rate is shown in Fig.~\ref{fig:lt-mean-msd}, where the long-term values decrease and saturate as the resetting rate increases beyond the corresponding critical points for convergence. Furthermore, the MSDs and means for different negative reallocation rates are depicted in Fig.~\ref{fig:mean-msd}. Negative reallocation with larger magnitude necessitates more frequent resetting to ensure convergence of both moments. In addition, as the reallocation strength becomes more negative, the long-term values of both the MSD and the mean decrease; however, the rate of this decrease slows as the reallocation further decreases. This observation is expected because more negative values of $\tau$ require higher resetting rates, which further constrain the space in which the agents can freely move, resulting in a lower long-term MSD. Consequently, this suggests that 'optimal' resetting is necessary to effectively balance growth and redistribution. If the resetting rate is too low, resources become concentrated, leading to high inequality. If it's too high, growth is hindered.

\begin{figure}[ht!]
\centering
\includegraphics[width=7cm]{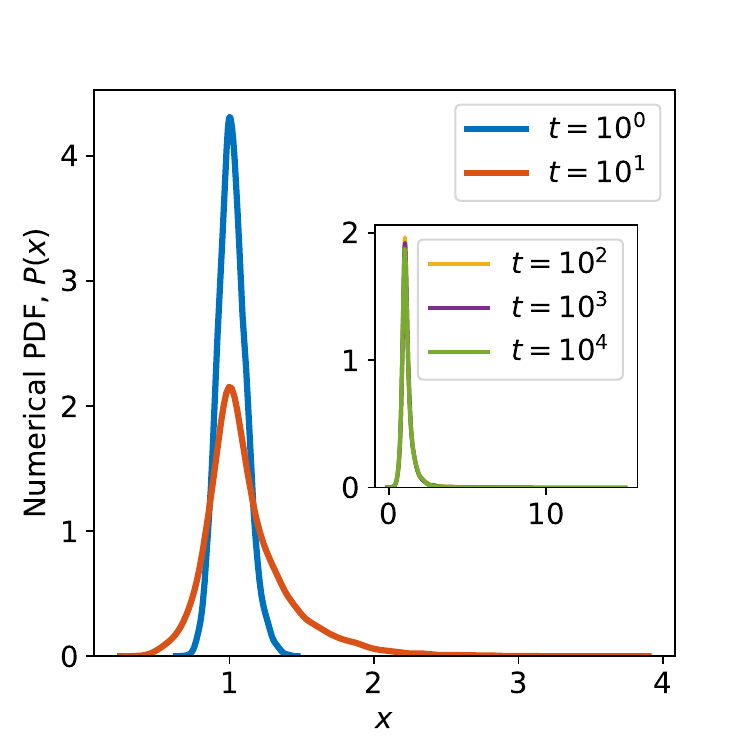}
\caption{\textbf{Numerical PDFs}. Parameters: $\mu=0.021, \sigma=\sqrt{0.01}, \tau = -0.01, r=0.15$ for different times.}
\label{fig:numerical-pdfs}
\end{figure}

As previously mentioned, a stationary probability distribution function does not exist in the negative reallocation regime. Thus, here we provide numerical evidence that resetting induces stationarity. In particular, we show in Fig.~\ref{fig:numerical-pdfs} the evolution of the PDF in the $\tau<0$ regime under resetting and see that the distribution stabilizes for $t\geq 10^2$. This fact is expected and supported by the observation that the second moment also becomes stationary at $t\approx 10^2$ (see right plot of Fig.~\ref{fig:mean-msd}).

\begin{figure}[ht!]
\centering
\includegraphics[width=7cm]{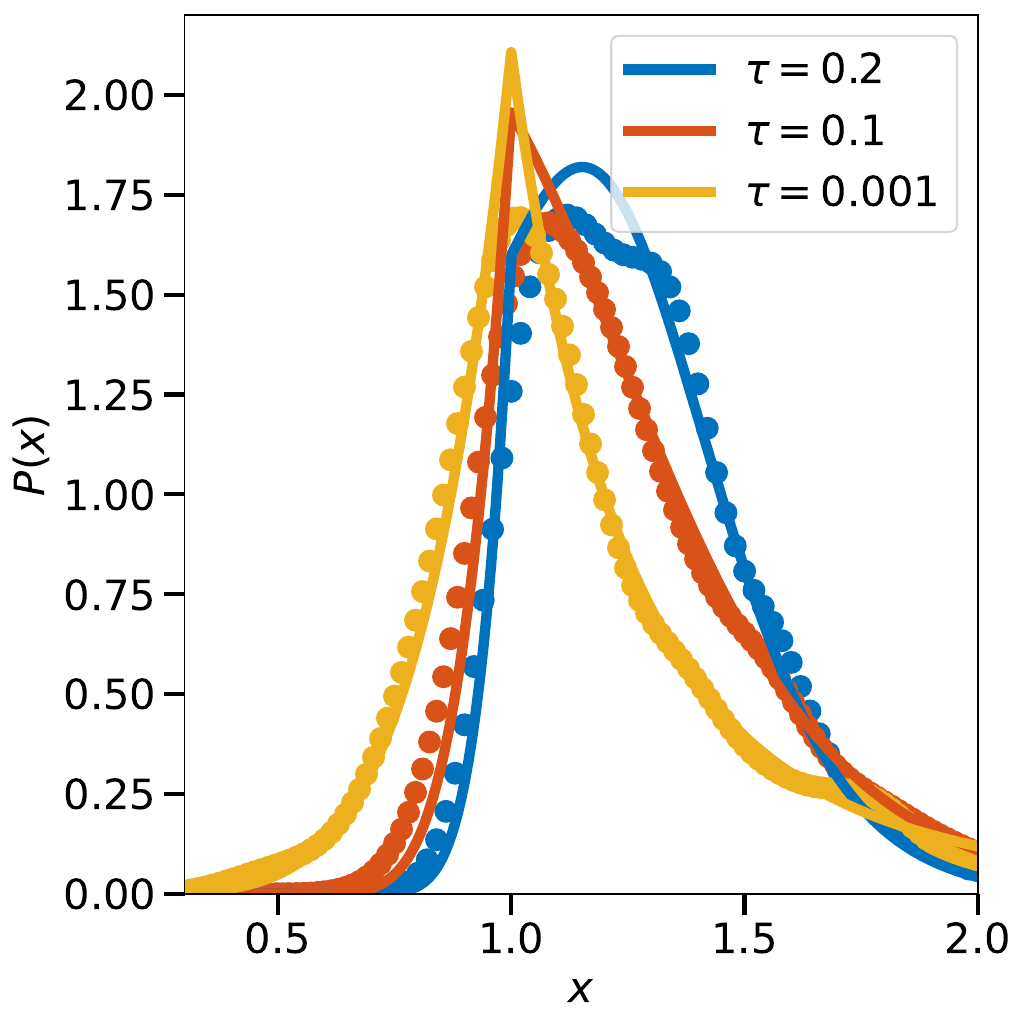}
\caption{Numerical solution of~(\ref{Fokker-Planck_approx_in third _regime}) (solid line) and simulation with process~(\ref{eq:rgbm_process}) (markers). Parameters: $\tau=0.05$, $\sigma=\sqrt{0.02}$, $\mu=0.021$, $r=0.17$}
\label{pdf_stationary}
\end{figure}

On the other hand, although the properties of the model for $\tau > 0$ are well known, we calculate the stationary PDF in this regime under resetting using our analytical approach. The PDFs for different rates of positive reallocation are shown in Fig.~\ref{pdf_stationary}.

Next, we explore the implications of the results from this section on wealth redistribution, focusing on how resetting influences mobility and wealth condensation.

\subsection{Mobility in the non-ergodic regime with resetting}

\begin{figure}[ht!]
\centering
\includegraphics[width=13cm]{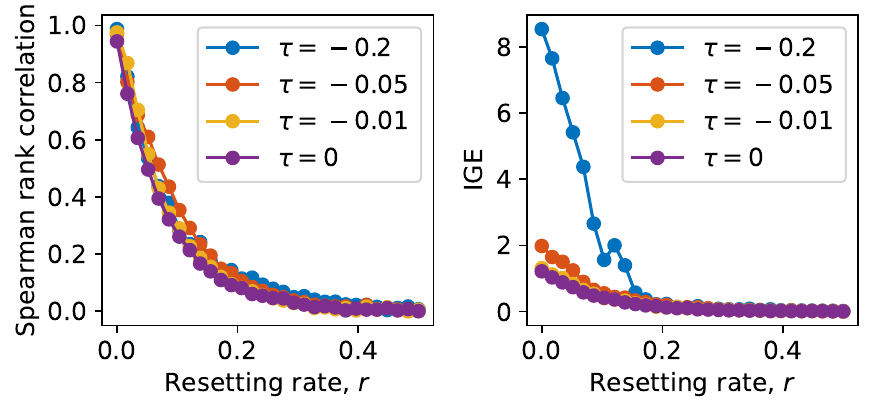}
\caption{\textbf{Mobility in the non-ergodic regime of RGBM}. Spearman's rank correlation as a function of the resetting rate for different values of $\tau$ (left). Intragenerational elasticity (IGE) as a function of the resetting rate and $\tau$ (right). The median results of $10^3$ simulations are shown. Parameters: $\mu=0.021, \sigma^2=0.02, \Delta=10, T=10^2, N=10^3$.}
\label{fig:mobility}
\end{figure}

Measures of economic mobility assess how individuals' wealth ranks change over time. When mobility is high, individuals have a greater likelihood of moving up or down in the wealth distribution over a certain period. On the other hand, when mobility is low, individuals are less likely to change their position in the wealth distribution. As mentioned in the introduction, we examine wealth mobility through Spearman's rank correlation and earnings elasticity in the $\tau<0$ regime (see Appendix~\ref{app:freezing-mobility}, paragraphs a. and b. for definitions of the mobility measures). It is known that the magnitude of the reallocation rate does not impact the extent of mobility \cite{stojkoski2024measures} (as measured by the previous two metrics) and that only the randomness ($\sigma$) in the system drives the changes in the observed wealth rankings. In both measures that we study, lower values indicate higher mobility. Here we recover the same result that the rank correlation is independent of the magnitude of the negative reallocation rate, and decreases as a function of the resetting rate in the same manner (see Fig~\ref{fig:mobility}). Furthermore, there exists some dependence between earnings elasticity and the strength of negative reallocation for smaller values of the resetting rate. As the resetting rate is further increased, earnings elasticity stabilizes around the same value regardless of $\tau$. We find that the rank correlation remains unaffected by the magnitude of the negative reallocation rate and decreases in the same way as a function of the resetting rate (see Fig~\ref{fig:mobility}).

\begin{figure}[ht!]
\centering
\includegraphics[width=8cm]{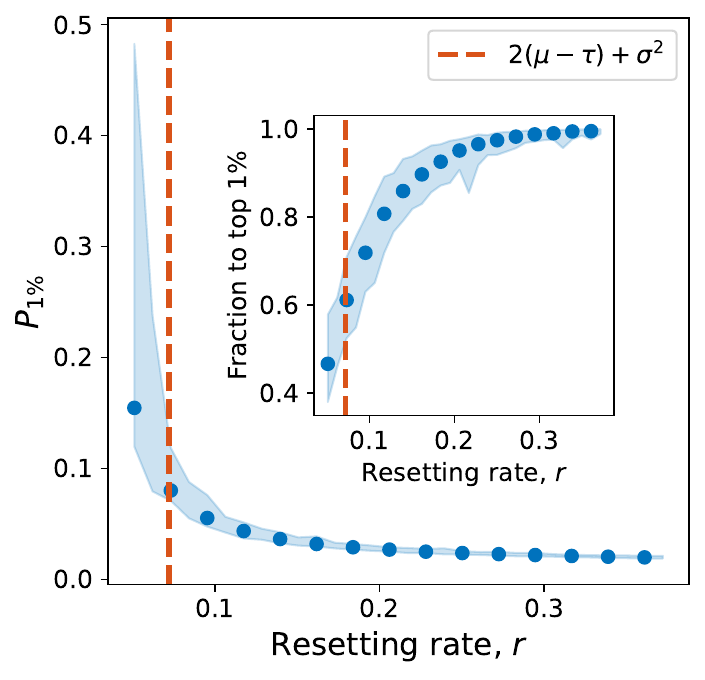}
\caption{\textbf{Degree of freezing}. Main plot shows the long time ($t=10^3$) probability of observing an agent that is among the largest 1\% as a function of the resetting rate. The median results are shown, and the filled region is the mininum and maximum of $10^3$ simulations. Inset plot shows the fraction of agents that reached the largest 1\%. The vertical orange dashed line marks the transition point $r_c$ at which the stable regime begins. These statistics are highly unstable in the first regime when $r<\mu-\tau$ and at the start of the second regime. Parameters: $\mu=0.021, \sigma^2=0.01, \tau=-0.01$.}
\label{fig:mobility_1}
\end{figure}

In addition, we study the degree of freezing with the probability $P_{1\%}(t)$ which shows how likely it is that the system is in one of the top 1\% of energy states at a given time $t$ (see Appendix~\ref{app:freezing-fraction}, paragraph a. for estimation details). In economic terms, this can be seen as a way to measure inequality. Concretely, as the resetting rate increases, the probability of observing extreme configurations decreases and stabilizes (see main plot in Fig.~\ref{fig:mobility_1}). Moreover, in the second regime ($\mu-\tau<r<2(\mu-\tau)+\sigma^2$; or before the vertical orange dashed line in the main plot in Fig.~\ref{fig:mobility_1}) the variance diverges and as a consequence there exists higher "wealth condensation" as indicated by the $P_{1\%}$ measure. After this transition point, the resetting is frequent enough for the variance to be finite; and the wealth of many agents stays close to the average. This observation is similar to the case of Bouchaud-Mézard model of wealth condensation~\cite{bouchaud2000wealth,ichinomiya2012bouchaud} where the critical point for variance convergence occurs when the positive coupling (reallocation), $J$, between agents on a network, exceeds some critical value $J_c$. Although, in our case $J=\tau$ is negative and the crucial point is that the transition to a non-condensation regime is induced by resetting, s.t. $r>r_c=2(\mu-\tau)+\sigma^2$. We also analyze the fraction of agents that reach the position of the top $1\%$ in the stationary distribution over a sufficiently long time frame, here $t=10^3$, (see Appendix~\ref{app:freezing-fraction}, paragraph b. for estimation details). For example, we observe that for larger resetting rates the fraction increases and stabilizes around 1 (see inset plot in Fig.~\ref{fig:mobility_1}). In other words, if the resetting rate is high enough (relative to other model parameters), then each individual will eventually experience the state of the top 1\% of the wealth distribution.

\section{Discussion} \label{sec:discussion}

The results of this paper offer a comprehensive understanding of how stochastic resetting alters the dynamics of the standard RGBM model, providing a potential pathway to address the challenges posed by non-ergodic and non-stationary processes. The introduction of stochastic resetting into the RGBM framework is a significant step forward, as it not only modifies the system's behavior but also offers solutions to stabilize and manage the redistribution of resources in fluctuating and unpredictable environments.

We identified three distinct regimes that the system can occupy, depending on the resetting rate. In the first regime, characterized by low resetting rate, the system behaves similar to the standard RGBM. This regime leads to uncertain outcomes that are difficult to predict and often resulting in concentration of resources, a phenomenon that mirrors the non-ergodic behavior observed in many natural and social systems, where wealth and resources tend to concentrate rather than distribute evenly. As the resetting rate further increases, the system transitions into a second regime, where the first moment, namely the mean value of the distribution becomes stationary. Here, we observed the initial stabilization effect of resetting, where the system starts to exhibit more predictable behavior. However, it is in the third regime, reached by further increasing the resetting rate, where the system achieves a higher level of stability, namely here both the first and the second moment (MSD) become stationary. This indicates that the system has reached a steady state, effectively mitigating the instability inherent in the RGBM model.

One of the most interesting aspects of our findings is the identification of a critical resetting rate at which the system's self averaging time becomes infinite. The analysis presented in Fig.~\ref{fig:critical_times} reveals a nuanced insight into the dynamics of our model. Specifically, it becomes clear that the size of the ensemble of individuals plays a less significant role in determining the critical resetting rate compared to the magnitude of the reallocation rate, $\tau$. It is the magnitude of $\tau$ that exerts a greater influence on the system's behavior. As $\tau$ becomes more negative, the redistribution dynamics become increasingly uneven and volatile, necessitating a higher frequency of resetting to achieve self-averaging. In other words, when resources are reallocated more aggressively, particularly in scenarios where wealth tends to concentrate the system requires more frequent interventions to maintain a more stable distribution. From Fig.~\ref{fig:lt-mean-msd} and Fig.~\ref{fig:mobility_1}, we can infer that while reaching a critical resetting rate is essential for achieving self-averaging, there are potential drawbacks if the resetting rate is increased much more beyond this critical threshold. Although a higher resetting rate may indeed lead to a more equitable distribution of wealth among agents, this comes at a cost: the overall wealth within the system is reduced. As a result, agents may find themselves with a more equal share of total wealth, but it becomes noticeably smaller as Fig.~\ref{fig:lt-mean-msd} suggests, limiting their ability to accumulate a substantial amount over time. This fact is further validated by the observed increase in the number of individuals who eventually reach the top 1$\%$ of the wealth distribution as the resetting rate increases, depicted in Fig.~\ref{fig:mobility_1}. However, as more individuals reach this part of the right tail, the time they spend there decreases accordingly. This finding underscores the critical importance of carefully calibrating the resetting rate in relation to the reallocation dynamics to ensure the system remains within the desired self-averaging regime.

In our exploration of mobility and inequality within the framework or RGBM, we found that higher resetting rates lead to reduced resources and wealth concentration and increased economic mobility. By using metrics such as Spearman's rank correlation and earnings elasticity we demonstrated that increased resetting results in a more evenly distributed flow of resources, with individuals having a greater likelihood of changing their position in the wealth distribution. This is particularly relevant in economic systems where wealth tends to concentrate.

Finally, we point out certain limitations of our work. The interpretation of stochastic resetting in an empirical setting is of crucial importance. In our case, it can be interpreted as an external (bankruptcy) event that resets an entity’s wealth. However, resetting stabilizes the first moment and as a result invalidates an important stylized fact that the mean wealth of an economy is a multiplicative growing quantity. Thus, using a different mechanism that stabilizes the non-ergodic regime and in the same time reproduces observed empirical facts is left for future work.

\section*{Acknowledgments}
The Authors acknowledge financial support by the German Science Foundation (DFG, Grant number ME 1535/12-1). This work is also supported by the Alliance of International Science Organizations (Project No. ANSO-CR-PP-2022-05). TS was supported by the Alexander von Humboldt Foundation. PT, LK and TS acknowledge support from the bilateral Macedonian-Austrian project - WTZ MK03/2024. PT acknowledge support from the Erwin Schr\"{o}dinger International Institute for Mathematics and Physics in Vienna. AP gratefully acknowledges the DST-SERB Start-up Research (Grant No. SRG/2022/000080) and the Department of Atomic Energy, India for research funding (via the “Modeling of Soft Materials”).

\appendix

\section{Properties of RGBM}\label{app:prop-rgbm}

The RGBM dynamics for the resources of entity $i$ at time $t$ are described by the following stochastic differential equation~\cite{marsili1998dynamical, bouchaud2000wealth, liu2017correlation},
\begin{equation} \label{eq:rgbm_process}
    d x_i (t) = x_i (t) (\mu dt + \sigma dW_i) - \tau (x_i(t) - \langle x \rangle_N) dt,
\end{equation}
where $\mu>0$ is the drift term, $\sigma$ is the noise amplitude, $dW_i$ is an independent Wiener increment, $\langle x \rangle_N $ is the ensemble or population average and $\tau$ is the rate of reallocation of wealth. The equation can be viewed as a combination of geometric Brownian motion (the first term) and mean-reverting term around the ensemble average, represented with Ornstein-Uhlenbeck process. In every time period $t$ everyone in the economy contributes a fraction $\tau$ of their wealth in the central pot. This is a case where all agents feel the same environment. It can be shown that the average wealth grows in time as $\exp{(\mu t)}$. Thus, it is more informative to consider the relative (rescaled) wealth $y_i = \frac{x_i}{\langle x \rangle_N}$ governed with the following stochastic equation:
\begin{equation}\label{eq:rgbm_normalized}
    dy_i(t) = y(t) \sigma dW_i - \tau (y(t)-1) dt,
\end{equation} 
and the corresponding Fokker-Planck equation:
\begin{equation}
    \frac{\partial}{\partial t}P(y,t) = \tau \frac{\partial}{\partial y} [(y-1)P(y,t)] + \frac{\sigma^2}{2} \frac{\partial^2}{\partial y^2} [y^2 P(y,t)].
\end{equation}

As previously mentioned, the reallocating geometric Brownian motion exhibits both ergodic and non-ergodic regimes, characterised by the sign of the reallocation parameter $\tau$:

\paragraph{Ergodic regime.} This regime occurs when the reallocation parameter is positive ($\tau > 0$). In this case wealths are positive with a Pareto tail around their mean value. This regime is characterized with reallocation of wealth from the richer to the poorer. Another property of this regime is the existence of a stationary distribution, and its form is that of an inverse gamma distribution
\begin{align}
    \mathcal{P}=\frac{(\zeta-1)^{\zeta}}{\Gamma (\zeta)}\exp{(-\frac{\zeta-1}{y})}y^{-(1+\zeta)},
\end{align}
where $\zeta=1+\frac{1+2 \tau}{\sigma^2}$ is the Pareto tail index, and can be used as a measure of economic equality~\cite{Cowell2011} and $\Gamma(\zeta)$ is the gamma function.

\paragraph{Non-ergodic regime.} On the other hand, when $\tau<0$ the model exhibits mean repulsion (the reallocation of wealth in this case is opposite from the previous regime and is from the poorer to the richer) and the population of trajectories splits into two groups, those above and below the mean. As a result, contrary to the ergodic regime, a stationary rescaled wealth distribution does not exist in this regime.

\section{Method of simulation}\label{app:method-simulation}

The main step to numerically simulate RGBM with resetting is to generate a trajectory using Eq.~\ref{stochastic_eq2}. Concretely, to obtain the distribution of the position of the particle at time $t$, we discretize the time $t = n\Delta t$, where $n$ is an integer. We initialize the position of the particle at $x(0)=1$, and then, at each step $(n = 1,2,3,\dots)$, the particle can either reset or it can evolve according to the laws of reallocating GBM. Thus,

\begin{enumerate}
    \item with probability $1-r\Delta t$ ($r$ is the rate of resetting); the particle undergoes reallocating GBM so that
\begin{align}\label{RGBM1}
    x(n \Delta t) =& \, x[(n -1)\Delta t] + x[(n-1)\Delta t ][\mu+\sigma \sqrt{\Delta t}\eta(n\Delta t)]
    -\tau[x[(n -1 )\Delta t ]-\langle x \rangle_N],
\end{align}
    where $\eta(n \Delta t)$ is a Gaussian random variable with mean 0 and variance 1, and $\Delta t$ is the microscopic time step;
    \item with complementary probability $r\Delta t$, resetting occurs such that
\begin{align}\label{RGBM2}
    x(n\Delta t) = x(0) = 1.
\end{align}
\end{enumerate}

A Python code for implementing the method of simulation can be found at: \url{https://github.com/pero-jolak/rgbm-resetting}.

\section{Method of estimating the degree of freezing and the fraction of individuals reaching the top $1\%$} \label{app:freezing-fraction}

\paragraph{Degree of freezing.}  We study the degree of freezing with the probability $P_{1\%}(t)$ which shows how likely it is that the system is in one of the top 1\% of energy states at a given time $t$. Numerically, this is done by reordering the trajectories, such that $x_1(n \Delta t) \geq x_2(n \Delta t) \geq \dots \geq x_N(n \Delta t)$, where $N$ is the number of trajectories and $n$ is an integer step $(n = 1,2,3,\dots)$. Then, $P_{1\%}(n \Delta t)$ in the period $n \Delta t$ is estimated as:
\begin{equation}
    P_{1\%}(n \Delta t) = \frac{\sum_{j}^{0.01N} x_j (n \Delta t)}{\sum_i^N x_i (n \Delta t)}
\end{equation}
A value of $P_{1\%}$ closer to 1 indicates a frozen configuration and consequently high concentration of wealth.

\paragraph{Fraction to top $1\%$.} The estimation procedure for the fraction of agents that reach the position of the top $1\%$ is as follows. We simulate the stochastic process described in Eqs.~\ref{RGBM1} and \ref{RGBM2} and generate $N=10^4$ trajectories of length $t=10^3$. Then, for each particle, we check if it has hit a predefined absorbing boundary, $x(n\Delta t)=y$, where $y$ represents the $99^{th}$ percentile of the distribution at $t=10^3$. Then, we calculate how many agents $k$ out of the total $N$ reached $y$. Finally, this procedure is repeated for each resetting rate, $r$.

\section{Definitions of mobility measures}\label{app:freezing-mobility}

In the following paragraphs we define the mobility measures analyzed in the main text: Spearman's rank correlation and Intragenerational earnings elasticity:

\paragraph{Spearman's rank correlation.} Spearman's rank correlation $\rho_{t,\Delta}$ is defined on a joint distribution of wealth at two points in time, $t$ and $t+\Delta$. Mathematically, it reads
\begin{align}
    \rho_{t,\Delta} = 1 - \frac{6\sum_i \left[rg\left(x_i\left(t\right)\right) - rg\left(x_i\left(t + \Delta \right)\right)\right]^2}{N\left(N^2-1\right)}\,,
\end{align}
where $rg(x)$ is the rank transformation of $x$. This measure is bounded between $-1$ and $1$. $\rho_{t,\Delta} = 1$ suggests perfect immobility, a state in which there is no change in wealth ranks between the two points in time. Lower values suggest greater wealth mobility.

\paragraph{Intragenerational earnings elasticity.} The earnings elasticity is defined as the slope $b_{t,\Delta}$ of the regression
\begin{align}
   \log\left(x_i\left(t+\Delta\right)\right) = b_0 + b_{t,\Delta} \log\left(x_i\left(t\right)\right) + u_i\,,
\end{align}
where $b_0$ is the intercept and $u_i$ is the error term. This is a simple linear regression and therefore,
\begin{align}
    b_{t,\Delta} = \mathrm{corr}\left[\log\left(x\left(t\right)\right),\log\left(x\left(t+\Delta\right)\right)\right] \frac{\mathrm{var}\left[\log\left(x\left(t+\Delta\right)\right)\right]}{\mathrm{var}\left[\log\left(x\left(t\right)\right)\right]}\,,
    \label{eq:iee-estimation}
\end{align}
where $\mathrm{corr}(x,y)$ is the correlation, between the variables $x$ and $y$, i.e., 
\begin{align}
    \mathrm{corr}\left[x, y\right] &= \frac{\mathrm{cov}[x, y]}{\sqrt{\mathrm{var}[x]} \sqrt{\mathrm{var}[y]}},
    \label{eq:autocorrelation}
\end{align}
with
\begin{align}
\mathrm{cov}[x, y] &\equiv \langle x y \rangle -\langle  x \rangle \langle y  \rangle 
\label{eq:bm_restart_covariance}
\end{align}
being the covariance of the same variables and $\mathrm{var}(x)$ is the variance of $x$. As with the rank correlation, lower EE also indicates greater mobility. However, this measure is unbounded and may take on any real values.

\bibliographystyle{unsrt}
\bibliography{bibliography}

\end{document}